\documentclass[conference]{IEEEtran}
\IEEEoverridecommandlockouts

\usepackage[utf8]{inputenc}
\usepackage[T1]{fontenc}
\usepackage{cite}
\usepackage{amsmath,amssymb,amsfonts}
\usepackage{graphicx}
\usepackage{textcomp}
\usepackage{xcolor}
\usepackage{url}
\usepackage[hidelinks]{hyperref}

\def\BibTeX{{\rm B\kern-.05em{\sc i\kern-.025em b}\kern-.08em
    T\kern-.1667em\lower.7ex\hbox{E}\kern-.125emX}}

\begin{document}

\title{A Tool for Anonymized Clinical Knowledge Retrieval and Speaker Interaction Analysis of Healthcare Tele-mentoring Sessions}

\author{
\IEEEauthorblockN{Parthaw Goswami}
\IEEEauthorblockA{\textit{University of Missouri}\\
Columbia, MO, USA\\
pgyn2@missouri.edu}
\and
\IEEEauthorblockN{Erfan Zareei}
\IEEEauthorblockA{\textit{University of Missouri}\\
Columbia, MO, USA\\
azzgg@missouri.edu}
\and
\IEEEauthorblockN{Amy Vogelsmeier}
\IEEEauthorblockA{\textit{University of Missouri}\\
Columbia, MO, USA\\
vogelsmeiera@missouri.edu}
\and
\IEEEauthorblockN{Dima Dandachi}
\IEEEauthorblockA{\textit{University of Missouri}\\
Columbia, MO, USA\\
dandachid@health.missouri.edu}
\and
\IEEEauthorblockN{Ben Casterline}
\IEEEauthorblockA{\textit{University of Missouri}\\
Columbia, MO, USA\\
casterlineb@health.missouri.edu}
\and
\IEEEauthorblockN{Mirna Becevic}
\IEEEauthorblockA{\textit{University of Missouri}\\
Columbia, MO, USA\\
becevicm@health.missouri.edu}
\and
\IEEEauthorblockN{Praveen Rao}
\IEEEauthorblockA{\textit{University of Missouri}\\
Columbia, MO, USA\\
praveen.rao@missouri.edu}
}

\maketitle

\begin{abstract}
We present a tool for anonymized retrieval of clinical knowledge and analysis of multi-speaker interactions in tele-mentoring sessions of Project ECHO (Extension for Community Healthcare Outcomes). Our tool's novel design includes: (a) a compound metadata filter that enforces strict speaker- and session-level isolation during retrieval, preventing cross-session context contamination; (b) MedGemma-4B as a locally-hosted language model for clinically grounded answer generation across three query modes; (c) interaction graphs to compute betweenness centrality, PageRank influence, and turn-taking patterns to quantify participation equity; and (d) a consistent alias-based anonymization scheme that prevents real participant names from appearing in any rendered output. During the demo, a healthcare provider (user) will interactively query multi-year ECHO sessions, explore participation dynamics through interaction graphs, compare sessions side-by-side, and assess answer quality via six reference-free RAG metrics, all without any cloud API dependency.
\end{abstract}


\section{Introduction}
Project ECHO (Extension for Community Healthcare Outcomes) is a globally adopted tele-mentoring model that connects specialists (hub members) with community-based healthcare providers through structured live-interactive sessions \cite{arora2017}. Each session follows a case-based learning format in which participants present de-identified patient cases, receive specialist guidance, and share didactic content, producing verbatim transcripts that capture clinical reasoning, peer-to-peer learning, and multidisciplinary discussion over time. With over 1600 active hubs operating across more than 200 countries, the corpus of ECHO transcripts represents a rich source of clinical education data for the study of discussion patterns, recurring clinical questions, and changes in participation over time \cite{unm_echo}.

Systematic analysis of ECHO transcripts at scale presents several challenges. Manual review is time-consuming and subjective; keyword search may miss clinically related concepts that are phrased differently across sessions; and general-purpose chatbots trained on public data risk generating plausible-sounding but clinically incorrect answers. Furthermore, participant privacy demands that speaker identities be protected before any material is shared beyond the research team. The inherently multi-speaker, session-structured nature of ECHO discourse introduces retrieval challenges that standard single-document retrieval augmented generation (RAG) pipelines \cite{lewis2020} are not designed to handle (e.g., without speaker- and session-aware filtering, a query directed at one participant may surface utterances from a different participant in a different session, undermining analytical validity). Beyond questions and answers (Q\&A), clinical education researchers require quantitative measures of participation equity, i.e., who speaks most, who bridges conversations, whether power gradients are narrowing over time, that are absent from existing tools.

To address these needs, we present ECHO-TD (Transcript Dashboard), a locally deployable, interactive Streamlit dashboard for analyzing ECHO telehealth session transcripts. ECHO-TD makes four contributions: (1) a session- and speaker-aware RAG pipeline using compound vector database metadata filtering to prevent cross-session contamination; (2) a graph-based participation analytics quantifying dominance, betweenness centrality, and turn-taking patterns; (3) three extended query modes (single-speaker, cross-speaker comparative, and multi-session longitudinal) addressing the full analytical question space of clinical education researchers; and (4) a reference-free evaluation framework of six RAG-aligned metrics plus a token-overlap hallucination-risk score, enabling self-contained evaluation without annotated reference answers.

\begin{figure*}[!t]
\centering
\includegraphics[width=0.80\textwidth]{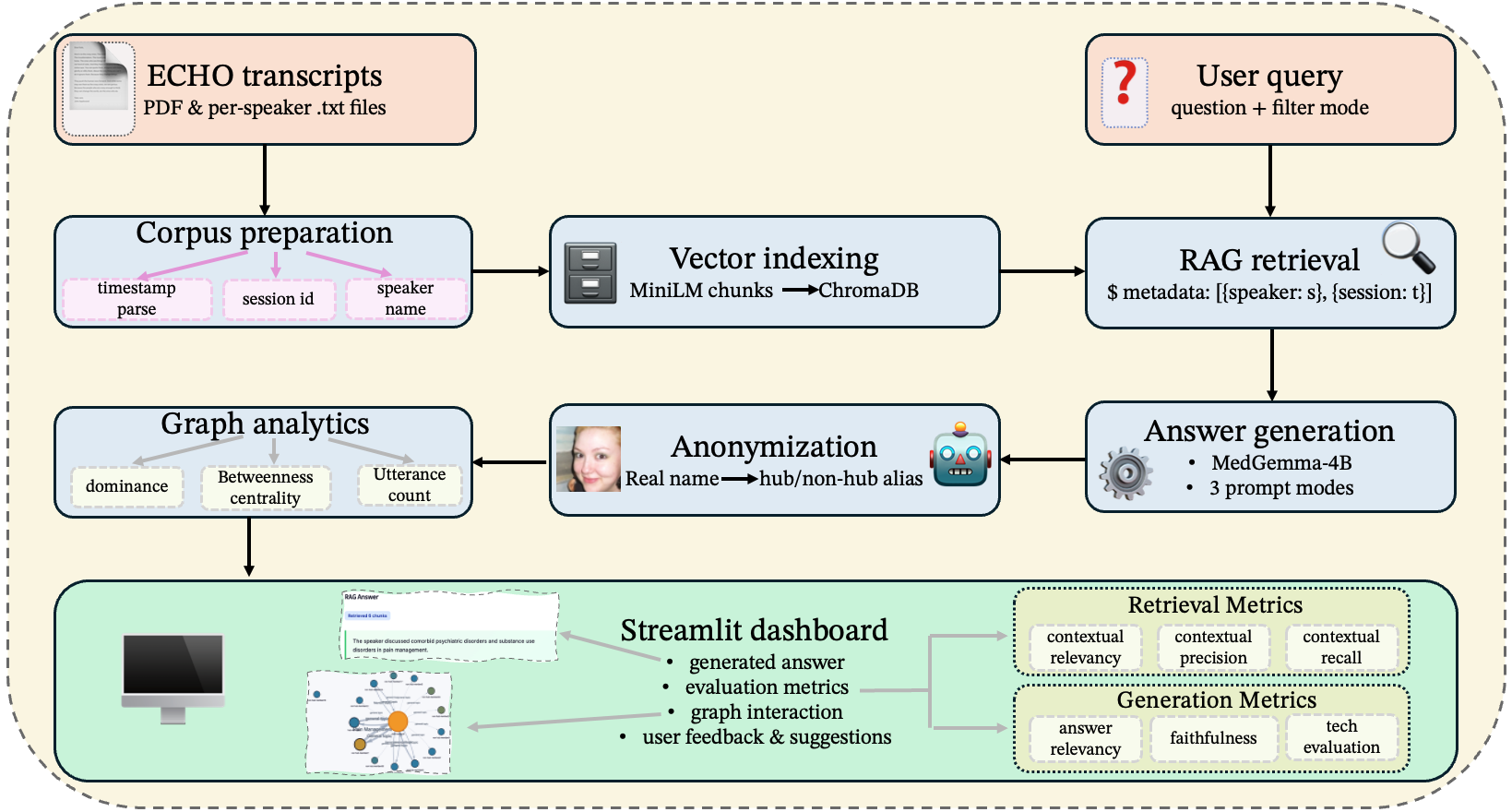}
\caption{System architecture of RAG-based ECHO telehealth sessions transcript analysis dashboard.}
\label{fig:arch}
\end{figure*}

The University of Missouri (MU) Institutional Review Board (IRB) acknowledged this project as Quality Improvement (QI) \#2132789 not requiring full board review. The demo video is available at: \href{https://mailmissouri-my.sharepoint.com/personal/pgyn2_umsystem_edu/_layouts/15/stream.aspx?id=%2Fpersonal%2Fpgyn2%5Fumsystem%5Fedu%2FDocuments%2FTranscript%20analysis%20demo%20video%2FDemo%20Video%20of%20transcript%20dashboard%20paper%20%28IEEE%20bigdata%29%2Emp4&ga=1&referrer=StreamWebApp%2EWeb&referrerScenario=AddressBarCopied%2Eview%2E1fb6c95a%2D6ef6%2D42aa%2D86b0%2Dad74f02762a7}{Demo Video}.

\section{Design of ECHO-TD}
The pipeline comprises six loosely coupled modules: corpus preparation, speaker-aware vector indexing, metadata-filtered RAG retrieval, answer generation, speaker anonymization, and interaction analytics. Operationally, ECHO-TD converts transcripts into speaker-labeled excerpts, stores them in a searchable local database, retrieves excerpts matching the selected speaker and session, generates a transcript-grounded answer, anonymizes speaker names, and summarizes interaction patterns. Figure~\ref{fig:arch} provides a conceptual schematic of the data flow.

\subsection{Corpus Preparation and Speaker-Aware Indexing}
Raw ECHO session transcripts are stored as PDF files. A custom line-level parser using the regular expression \texttt{[YYYY.MM.DD HH:MM:SS]} reconstructs per-speaker utterance streams, tagging each utterance with its \texttt{session\_id} (the date component of the timestamp) and the speaker's normalized name. Ten sessions are processed: five from early 2016 (February--June) and five from early 2020 (January--March), covering chronic pain management, opioid prescribing protocols, and psychiatric comorbidities. A pre-processing validation step checks that each parsed utterance carries a valid timestamp and a non-empty speaker name and detects split utterances broken across multiple lines.

The first key design feature of ECHO-TD is the speaker-aware vector index. Each utterance block is split into overlapping 512-token chunks with a 64-token stride, encoded with sentence-transformers/all-MiniLM-L6-v2 \cite{reimers2019} (producing 384-dimensional dense vectors) and stored in a ChromaDB persistent collection with two mandatory metadata fields: \texttt{speaker} (normalized real name) and \texttt{session\_id} (YYYY.MM.DD date string). After full indexing, the collection holds approximately 3,200 chunks. The indexing pipeline is idempotent: existing chunks are skipped rather than duplicated, allowing re-runs after adding new sessions without corrupting the collection.

\subsection{Metadata-Filtered RAG Retrieval}
The second key design feature is the compound metadata filter. Given query $q$, speaker $s$, and session $t$, ECHO-TD constructs a ChromaDB where clause: \texttt{\{"speaker": $s$, "session\_id": $t$\}}. Standard vector similarity search ranks by embedding proximity alone and retrieves semantically similar content from any speaker or session in the corpus; the compound filter reduces the candidate set to the exact (speaker, session) pair, achieving zero cross-session chunk leakage across the queries. Retrieved chunks (top-$k$, user-configurable 2--20) are assembled into a structured context string with labelled headers \texttt{[chunk N | speaker=<alias> | session=<id>]} to orient the language model and enable chunk-level attribution in the evaluation metrics. The first query in a session incurs a one-time ChromaDB collection load ($\sim$3.2~s), after which the collection is cached in Streamlit session state. Subsequent retrievals complete in under 0.5~s.

\begin{figure*}[!t]
\centering
\includegraphics[width=0.85\textwidth]{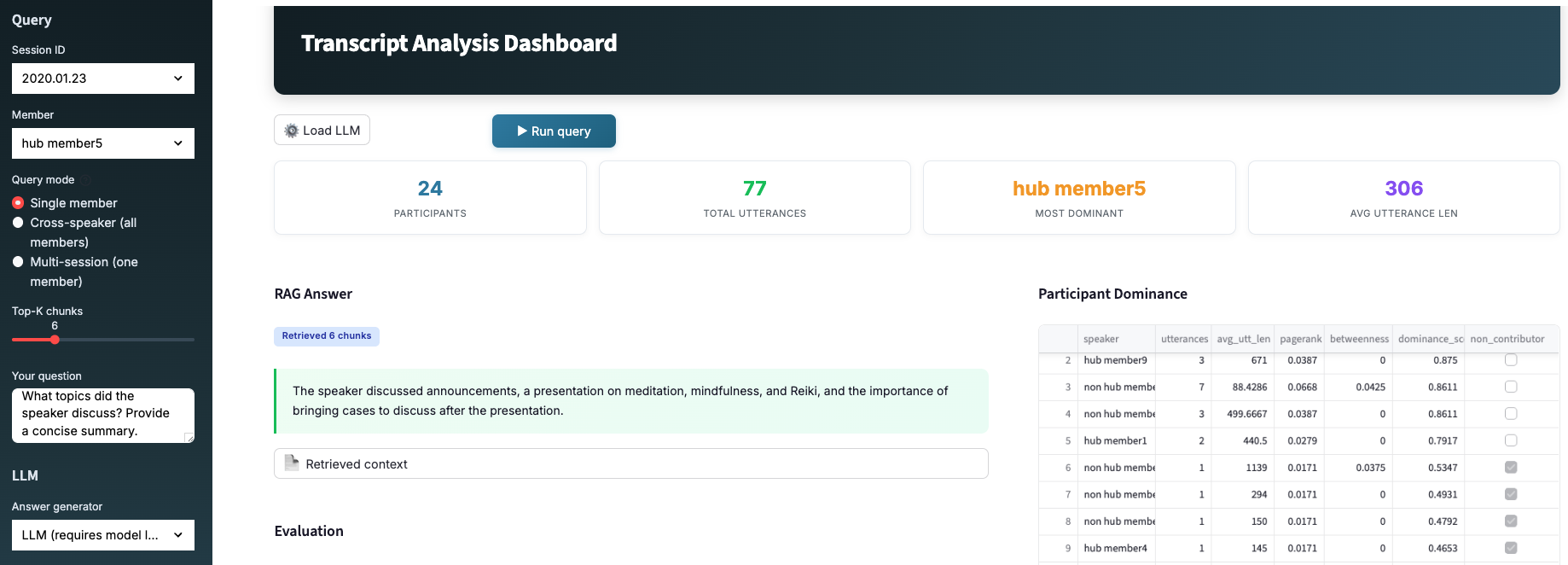}
\caption{Transcript analysis dashboard showing speaker names are replaced by hub/non-hub members.}
\label{fig:dashboard}
\end{figure*}

\subsection{Answer Generation and Speaker Anonymization}
Retrieved chunks are passed to MedGemma-4B \cite{sellergren2025}, a 4-billion parameter multimodal language model fine-tuned on medical data, running locally on CPU in float32 precision via the HuggingFace Transformers pipeline. All prompts include an explicit instruction to abstain when the answer is not in the provided context, suppressing confabulation. Three specialized prompt templates handle the three query modes. \textbf{Single-speaker mode} retrieves top-$k$ chunks for a single (speaker, session) pair and generates a context-grounded answer. \textbf{Cross-speaker comparative mode} retrieves $\lfloor k/N \rfloor$ chunks from each of the $N$ speakers present in the selected session, assembles a unified context grouped by speaker alias, and uses a contrastive prompt to identify points of agreement, disagreement, and the speaker who held the most unique view (directly addressing questions such as ``Who disagreed about the opioid dosing protocol?''). \textbf{Multi-session longitudinal mode} retrieves $\lfloor k/S \rfloor$ chunks per session for a selected speaker and uses a synthesis prompt to summarize per-session contributions, highlight evolution over time, and flag sessions where the topic was not addressed (tracking individual clinician development across a multi-year ECHO cohort).

All speakers are classified as hub members or non-hub members based on their work activity role and assigned consistent aliases (\texttt{hub\_member1}, \texttt{non\_hub\_member1}, \dots) applied at render time to the answer text, context chunk headers, participant dominance table, and graph node labels, so that no real name appears in any rendered, exported, or screenshot-captured output (Figure~\ref{fig:dashboard}).

\subsection{Interaction Graph Analytics}
The third key design feature is the graph-based participation analytics. A directed NetworkX DiGraph $G=(V,E)$ is constructed for each session by iterating over chronologically ordered utterances and adding a weighted edge $A \rightarrow B$ each time speaker $B$ immediately follows speaker $A$. Edge weight equals the number of such consecutive transitions (turn-taking frequency). Three centrality metrics are computed: (i) normalized betweenness centrality (speaker acts as a bridge between others) using Brandes' $O(nm)$ algorithm \cite{brandes2001}; (ii) PageRank \cite{page1999} (conversational influence) with border thickness of nodes; and (iii) a composite dominance score defined as the mean of the percentile ranks of utterance count, character count, and PageRank across all nodes in the session. Speakers at or below the 15th percentile of utterance count are flagged as non-contributors, providing an early-warning signal for facilitators. Figure~\ref{fig:graph} shows the graph visual interactions of a selected speaker with other speakers in a selected ECHO session. The graph is rendered as a physics-simulated, draggable HTML widget via pyvis 0.3.2. Node color encodes betweenness centrality on a blue$\rightarrow$amber$\rightarrow$red ramp; node size encodes utterance count (log-scaled); and edge width encodes turn-taking frequency. Hovering over any node reveals a plain-English tooltip listing numeric metrics with natural-language interpretations.

\textbf{Session comparison mode} renders two side-by-side pyvis graphs for any two user-selected sessions, sharing the same color scale and node-sizing rules to enable direct visual comparison. A delta table below the graphs quantifies per-speaker changes in utterance count, betweenness centrality, and dominance score between the two sessions.

\subsection{Evaluation Metrics}
Es et al.\ \cite{es2024} proposed RAGAS (Retrieval Augmented Generation Assessment), a framework of reference-free metrics for evaluating RAG outputs using the same embedding model as retrieval. We adopt reference-free proxies aligned with the RAGAS framework and six metrics are reported for every generated answer (Figure~\ref{fig:metrics}). Contextual relevancy (the maximum cosine similarity between the query embedding and any retrieved chunk embedding), contextual precision (the mean cosine similarity between the answer embedding and the top-3 chunk embeddings), and contextual recall (the maximum cosine similarity between the answer embedding and any retrieved chunk embedding) capture retrieval quality. These metrics use the same all-MiniLM-L6-v2 model for embedding, to ensure alignment with the retrieval representation space. Answer relevancy is the cosine similarity between query and answer embeddings, measuring on-topic responsiveness. Faithfulness combines semantic and lexical grounding ($0.6 \times \max$ answer-to-chunk embedding similarity $+\ 0.4 \times$ Jaccard token overlap). The technical-term heuristic measures the proportion of predefined clinical keywords present in the answer. Hallucination risk occurs when the model generated answer is not supported by the provided context. We estimate hallucination risk using the proportion of meaningful words in the generated answer that are not supported by the retrieved context after removing common stop words. Risk bands: low ($h<0.3$), medium ($0.3$--$0.6$), high ($h>0.6$).

\subsection{Dashboard Usability Feedback}
\begin{figure}[!t]
\centering
\includegraphics[width=\columnwidth]{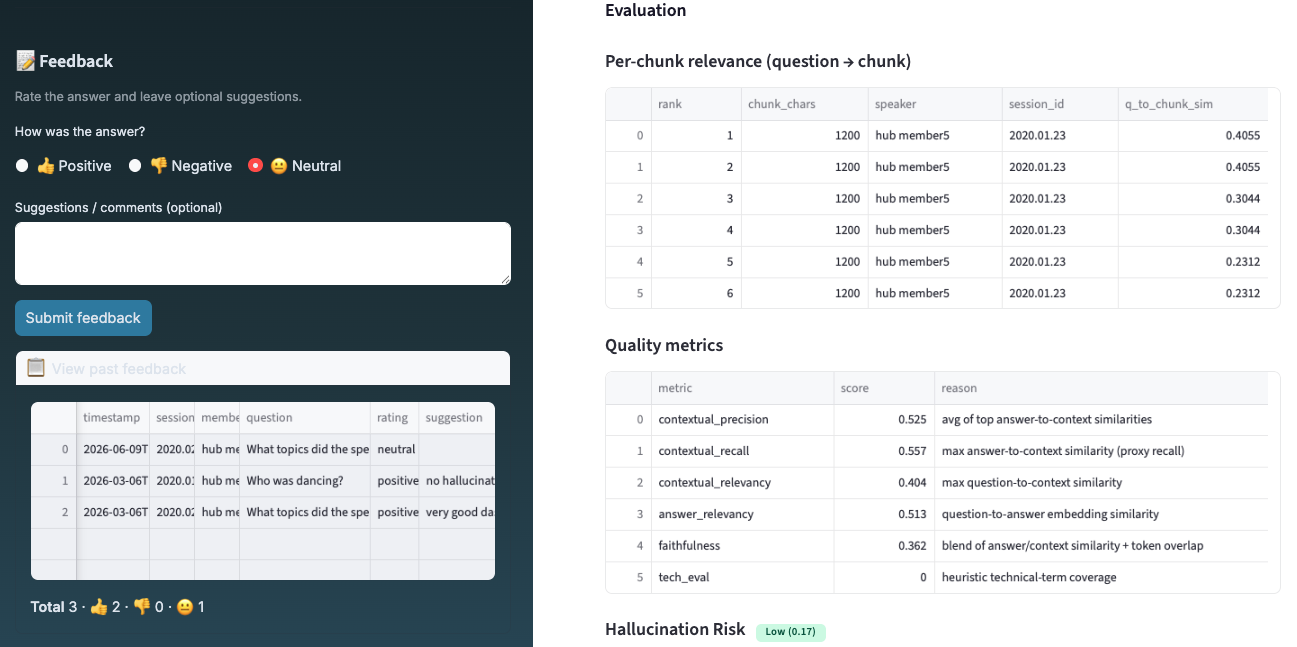}
\caption{Dashboard showing reference free evaluation metrics of RAG-based answer for one of the user queries and feedback section in the sidebar.}
\label{fig:metrics}
\end{figure}

The dashboard sidebar includes a feedback section where healthcare providers (users) can rate responses, submit suggestions, browse past feedback provided by other users, and explore a history of questions asked by past users and reviewers, which serves as a helpful reference for discovering the types of queries the dashboard is best equipped to handle, ultimately guiding new users toward more effective and informed interactions with the system. Figure~\ref{fig:metrics} shows the feedback section and previous questions/suggestions exploration in the dashboard.

\section{Demonstration Scenarios}
ECHO-TD was implemented in Python using Streamlit, ChromaDB, LangChain, sentence-transformers, NetworkX, and pyvis. All processing runs locally. A user will interactively experience four scenarios (Figure~\ref{fig:dashboard}).

\textbf{Scenario 1: Single-Speaker Q\&A.} The user selects a session and a single speaker alias from the sidebar. After entering a natural-language question (e.g., \emph{``How did this member approach the case of chronic opioid dependence?''}), ECHO-TD retrieves the top-$k$ chunks filtered to that exact pair (speaker, session) and generates a clinically grounded answer via MedGemma-4B. The user will observe the six reference-free evaluation metrics, the hallucination-risk score, and the per-chunk cosine-similarity table, which allows independent assessment of retrieval quality (Figure~\ref{fig:metrics}). The user can confirm that the answer contains no real speaker names, only hub/non-hub aliases.

\textbf{Scenario 2: Cross-Speaker Comparative Q\&A.} The user can ask \emph{``Which members discussed psychiatric comorbidities from different clinical standpoints?''}, ECHO-TD retrieves chunks from all $N$ speakers in the selected session, assembles a unified context grouped by alias, and returns a contrastive answer identifying agreement, disagreement, and the speaker with the most unique view. A per-member chunk breakdown table shows retrieved chunk counts and mean cosine similarity, highlighting speakers who contributed little retrievable content to the topic.

\textbf{Scenario 3: Longitudinal Analysis.} The user selects a single speaker and poses a topic-level query across all sessions that speaker attended. ECHO-TD assembles context labelled by session date, applies the longitudinal synthesis prompt, and returns a chronological thread summarizing how the speaker's position on the query topic evolved over time, noting sessions where the topic was absent. The user will observe that longitudinal mode carries a longer latency than single-speaker mode.

\begin{figure}[!t]
\centering
\includegraphics[width=\columnwidth]{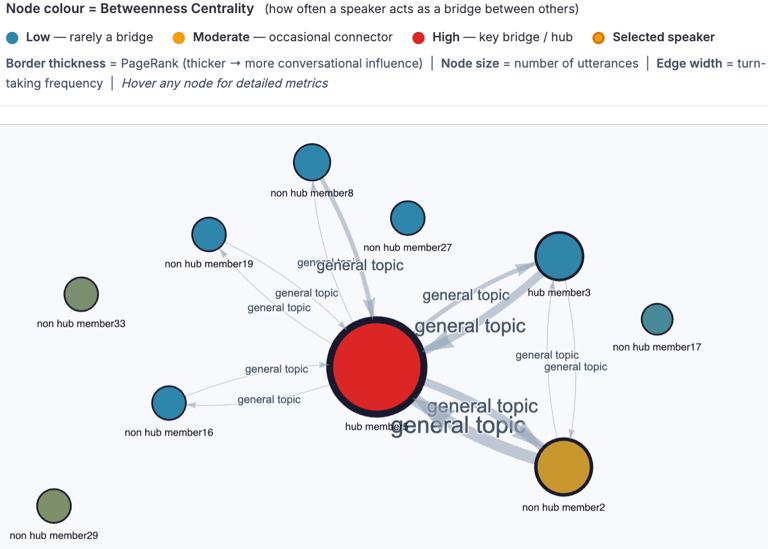}
\caption{Dashboard showing interactions (betweenness centrality, conversational influence) of the selected speaker with other speakers in a selected session.}
\label{fig:graph}
\end{figure}

\textbf{Scenario 4: Interaction Graph Exploration.} The user selects a speaker and a session to view the interaction graph (Figure~\ref{fig:graph}). Node color, size, and edge width encode betweenness centrality, utterance count, and turn-taking frequency, respectively. Next, the user checks session comparison to render two pyvis graphs side-by-side and reads the delta table to identify speakers whose participation shifted notably.

\section*{Acknowledgment}
This presentation was supported by the Health Resources and Services Administration (HRSA) of the U.S. Department of Health and Human Services (HHS) under the Office of the Advancement of Telehealth (OAT) cooperative agreement number U3G55162, which has a total award amount of \$775,000 for the approved project period. The contents are those of the author(s) and do not necessarily represent the official views of, nor an endorsement by, HRSA, HHS, or the U.S. Government.

\bibliographystyle{IEEEtran}
\bibliography{references}

\end{document}